# A Physics-Guided Modular Deep-Learning Based Automated Framework for Tumor Segmentation in PET Images

Kevin H. Leung[1,2], Wael Marashdeh[3], Rick Wray[4], Saeed Ashrafinia[2,5], Martin G. Pomper[1,2], Arman Rahmim[6] and Abhinav K. Jha[7]

[1] Department of Biomedical Engineering, Johns Hopkins University, Baltimore, MD, USA  
[2] The Russell H. Morgan Department of Radiology, Johns Hopkins University, Baltimore, MD, USA  
[3] Department of Radiology and Nuclear Medicine, Jordan University of Science and Technology, Arramtha, Jordan  
[4] Memorial Sloan Kettering Cancer Center, Greater New York City Area, NY, USA  
[5] Department of Electrical & Computer Engineering, Johns Hopkins University, Baltimore, MD, USA  
[6] Departments of Radiology and Physics, University of British Columbia, Vancouver, BC, Canada  
[7] Department of Biomedical Engineering and Mallinckrodt Institute of Radiology, Washington University in St. Louis, St. Louis, MO, USA

E-mail: a.jha@wustl.edu

**Abstract**

An important need exists for reliable PET tumor-segmentation methods for tasks such as PET-based radiation-therapy planning and reliable quantification of volumetric and radiomic features. The purpose of this study was to develop an automated physics-guided deep-learning-based PET tumor-segmentation framework that addresses the challenges of limited spatial resolution, high image noise, and lack of clinical training data with ground-truth tumor boundaries in PET imaging. We propose a three-module PET-segmentation framework in the context of segmenting primary tumors in 3D $^{18}$F-fluorodeoxyglucose (FDG)-PET images of patients with lung cancer on a per-slice basis. The first module generates PET images containing highly realistic tumors with known ground-truth using a new stochastic and physics-based approach, addressing lack of training data. The second module trains a modified U-net using these images, helping it learn the tumor-segmentation task. The third module fine-tunes this network using a small-sized clinical dataset with radiologist-defined delineations as surrogate ground-truth, helping the framework learn features potentially missed in simulated tumors. The framework's accuracy, generalizability to different scanners, sensitivity to partial volume effects (PVEs) and efficacy in reducing the number of training images were quantitatively evaluated using Dice similarity coefficient (DSC) and several other metrics. The framework yielded reliable performance in both simulated (DSC: 0.87 (95% CI: 0.86, 0.88)) and patient images (DSC: 0.73 (95% CI: 0.71, 0.76)), outperformed several widely used semi-automated approaches, accurately segmented relatively small tumors (smallest segmented cross-section was 1.83 cm$^2$), generalized across five PET scanners (DSC: 0.74 (95% CI: 0.71, 0.76)), was relatively unaffected by PVEs, and required low training data (training with data from even 30 patients yielded DSC of 0.70 (95% CI: 0.68, 0.71)). A modular deep-learning-based framework yielded reliable automated tumor delineation in FDG-PET images of patients with lung cancer using a small-sized clinical training dataset, generalized across scanners, and demonstrated ability to segment small tumors.

## 1. Introduction

Accurate tumor delineation in positron emission tomography (PET) is important for many clinical tasks, such as PET-based radiation-therapy planning and reliable quantification of volumetric and radiomic features (Jha et al 2017, Foster et al 2014, Mena et al 2017a). However, PET tumor delineation is challenging due to the relatively poor spatial resolution and high noise of PET images (Foster et al 2014). Typically, tumors in PET images are segmented manually, which is tedious, time-consuming, expensive (Foster et al 2014, Bagci et al 2013), and suffers from substantial inter- and intra-reader variability (Shah et al 2012, Giraud et al 2002). Further, accurate tumor boundaries are hard to obtain manually due to partial volume effects (PVEs) arising due to low resolution in PET images. To address these issues, several computer-aided segmentation methods for PET images have been developed (Foster et al 2014). Common methods include approaches based on thresholding (Foster et al 2014, Mena et al 2017a, Shah et al 2012, Mena et al 2017b, Sridhar et al 2014), gradient-detection (Werner-Wasik et al 2012), and using image-data statistics or assuming prior knowledge about the tumor (Foster et al 2014, Aristophanous et al 2007, Hatt et al 2009, Jha et al 2010, Soufi et al 2016, Layer et al 2015, Belhassen and Zaidi 2010). Those methods have demonstrated promise but suffer from limitations, such as requiring manual input (e.g. tumor seed pixel or region of interest (ROI) around the tumor) (Foster et al 2014, Mena et al 2017b, Sridhar et al 2014, Werner-Wasik et al 2012, Aristophanous et al 2007, Hatt et al 2009), sensitivity to PVEs (Foster et al 2014, Brambilla et al 2008), limitations when assumptions are not satisfied (Belhassen and Zaidi 2010), and need for recalibration with different scanners (Zaidi et al 2012). Given these limitations, there is an important need to develop more accurate, generalizable, robust, and automated PET segmentation methods.

Our objective in this paper was to develop a method that, when given a PET image slice with a tumor, automatically localizes and accurately delineates the tumor. In this context, deep-learning (DL)-based methods, in particular those based on convolutional neural networks, such as U-net, have shown substantial promise in medical-image segmentation – especially in anatomical imaging modalities such as computed tomography (CT) and magnetic resonance imaging (MRI) (Litjens et al 2017, Ronneberger et al 2015, Pereira et al 2016). While some recent studies explored DL-based methods for PET segmentation (Blanc-Durand et al 2018, Zhao et al 2018), several challenges remain to be addressed (Foster et al 2014, Litjens et al 2017). PET images have low spatial resolution and high image noise compared to anatomical imaging, which makes segmentation challenging (Foster et al 2014) and complicates the task of determining ground-truth for training DL-based approaches. DL-based segmentation methods typically use manual delineation as surrogate ground-truth which, in the context of PET images, has several issues as mentioned above. These include the important issues of limited accuracy and high variability, which limit the use of manual segmentations as ground truth (Foster et al 2014, Shah et al 2012, Giraud et al 2002). Further, for effective training, DL-based methods typically require large training sets (order of thousands), which are not readily available since PET is a relatively low-volume modality (Shen et al 2017).

To address these challenges, instead of training a conventional DL-segmentation approach only on limited clinical manually delineated training images, we propose a physics-guided three-module DL framework (Fig. 1). The framework was developed and comprehensively evaluated in the context of segmenting primary tumors in $^{18}$F-fluorodeoxyglucose (FDG)-PET images of patients with lung cancer and yielded accurate segmentation with a small clinical training set, generalized across different scanners, and was relatively insensitive to PVEs.

## 2. Materials and methods

This was an IRB approved, HIPAA-compliant, retrospective study, with a waiver for obtaining informed consent. Data from 160 patients (91 Male, 69 Female, mean age 63.2±11.7 [standard deviation] years; range, 27–90 years) with biopsy-proven lung cancer and a measurable pulmonary tumor on staging $^{18}$F-FDG PET/CT was used. Patients with a second primary malignancy were excluded. Detailed patient characteristics are given in the Appendix (Table A1). Standard imaging protocol involved FDG administration of 0.22 mCi/kg and image acquisition 60 minutes post-injection. The data was acquired across five different scanners: Discovery LS (N=104), Discovery RX (N=40), Discovery HR (N=7), Discovery ST (N=7), and Discovery STE (N=2). Details on scanner and reconstruction parameters are in Table 1.

*2.1 The modular U-net based DL framework*

A modular DL-based framework was developed in the context of segmenting 3D PET images on a per-slice basis. The framework, when input a PET image slice containing a tumor, localized and segmented the tumor (Fig. 1). At its core, the framework had a modified U-net (mU-net) (U-net with modifications detailed in Section 2.1.2) (Fig. 2). Training DL approaches require large datasets with accurate ground-truth (Shen *et al* 2017). Typically, only a small number of PET clinical images with only surrogate ground-truths (manual delineations) are available. In the DL literature, the use of simulated data to train DL methods has demonstrated promise (Creswell *et al* 2018, Gong *et al* 2018) and motivated our approach to use realistic simulations that model the PET imaging physics to address training-data scarcity. However, simulated tumors may not be fully representative of tumors from the patient population and may not incorporate all tumor features. To address these issues, we developed a three-module framework (Fig. 1). The first module simulates a large number of images containing realistic 2D tumors with known ground-truth addressing the issue of lack of training data. Using a new stochastic kernel-density estimation

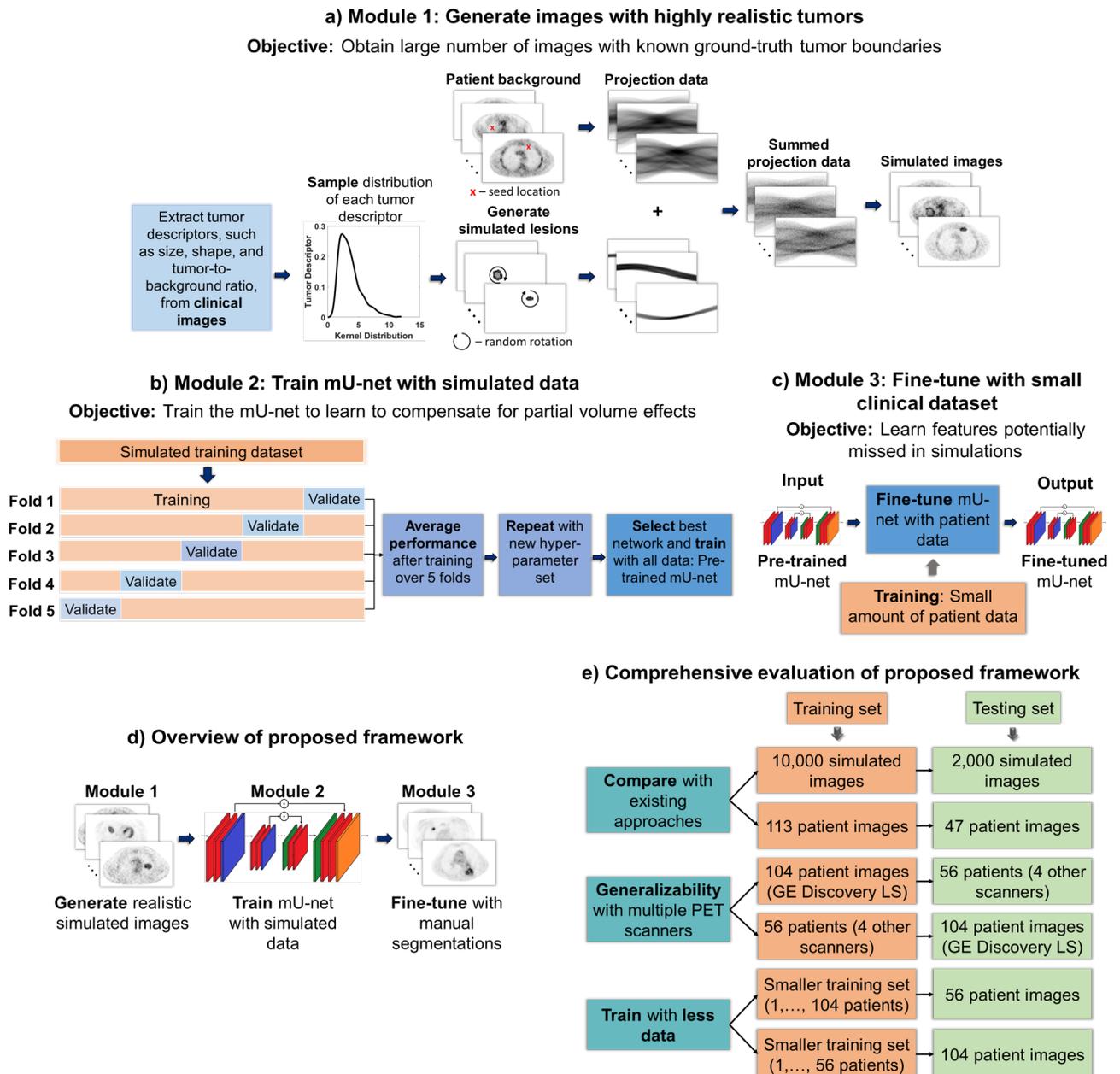

**Figure 1**: Illustration of the generation of simulated $^{18}$F-fluorodeoxyglucose (FDG)-PET images (a), workflow for the 5-fold cross-validation process used to optimize and train the modified U-net (mU-net) with simulated data (b), fine-tuning the mU-net with patient data (c), overview of the proposed modular framework (d), and evaluation of the proposed framework (e).

**Table 1**: Technical acquisition and reconstruction parameters of PET/CT systems.

| Parameter | Discovery LS | Discovery RX | Discovery HR | Discovery ST | Discovery STE |
|---|---|---|---|---|---|
| PET transaxial FOV (mm) | 550 | 700 | 700 | 700 | 700 |
| PET axial FOV (mm) | 153 | 153 | 157 | 157 | 153 |
| Reconstruction method | OSEM | OSEM | OSEM | OSEM | OSEM |
| Subsets | 28 | 21 | 21 | 21 | 20 |
| Iterations | 2 | 2 | 2 | 2 | 2 |
| Randoms correction method | RTSUB | SING | DLYD | DLYD | DLYD |
| Attenuation correction method | CT | CT | None | CT | CT |
| Scatter correction method | Convolution subtraction | Convolution subtraction | None | Convolution subtraction | Convolution subtraction |
| Energy window (keV) | 300 – 650 | 425 – 650 | 425 – 650 | 375 – 650 | 375 – 650 |
| Voxel size (mm$^3$) | 3.91 × 3.91 × 4.25 | 4.69 × 4.69 × 3.27 | 4.69 × 4.69 × 3.27 | 5.47 × 5.47 × 3.27 | 5.47 × 5.47 × 3.27 |

CT: computed tomography, DLYD: delayed event subtraction method, FOV: field of view, OSEM: ordered subset expectation-maximization, PET: positron emission tomography, RTSUB: real-time delayed event subtraction method, SING: singles-based correction.

(KDE) and a physics-based approach, this module generates realistic 2D tumors using tumor descriptors obtained from clinical data to capture the observed variability in actual populations. Further, the module uses PET physics with the intent of accounting for partial volume effects in the segmentation process. The second module trains and optimizes the mU-net using these simulated images such that the mU-net learns the tumor-segmentation task for low-resolution images. The third module fine-tunes the mU-net with patient data to learn tumor features missed in simulated tumors. The modules are further described below:

*2.1.1 Module 1 – Stochastic and physics-guided generation of realistic tumors.*   A novel KDE and physics-based approach was developed to simulate FDG-PET image slices with realistic tumors (Fig. 1a). Since segmentation performance is especially influenced by tumor shapes and size, count statistics, and tumor-to-background ratio, it was important to accurately account for these parameters when simulating the tumors. For this purpose, we obtained the distributions of these parameters from clinical data and used these distributions to simulate the realistic tumors. Tumor descriptors were extracted from clinical FDG-PET images, including first- and second-order statistics for the intensity, size, shape, intra-tumor heterogeneity, and tumor-to-background intensity ratio. The background intensities were obtained from non-tumor pixels present within a circular ROI around the tumor. Tumor shape was quantified using five harmonic elliptical Fourier shape descriptors (Kuhl and Giardina 1982). Tumor size was quantified using diameter and volume. Each metric was modeled with a kernel distribution.

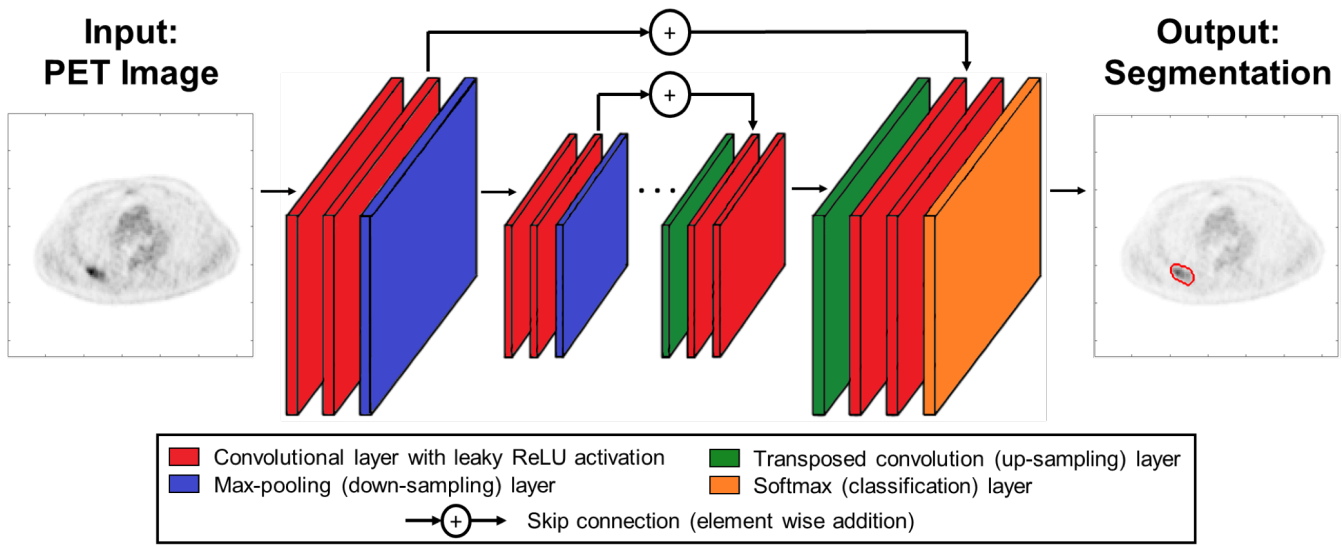

**Figure 2**: Illustration of the mU-net architecture present in the second module of the proposed framework. ReLU: rectified linear unit.

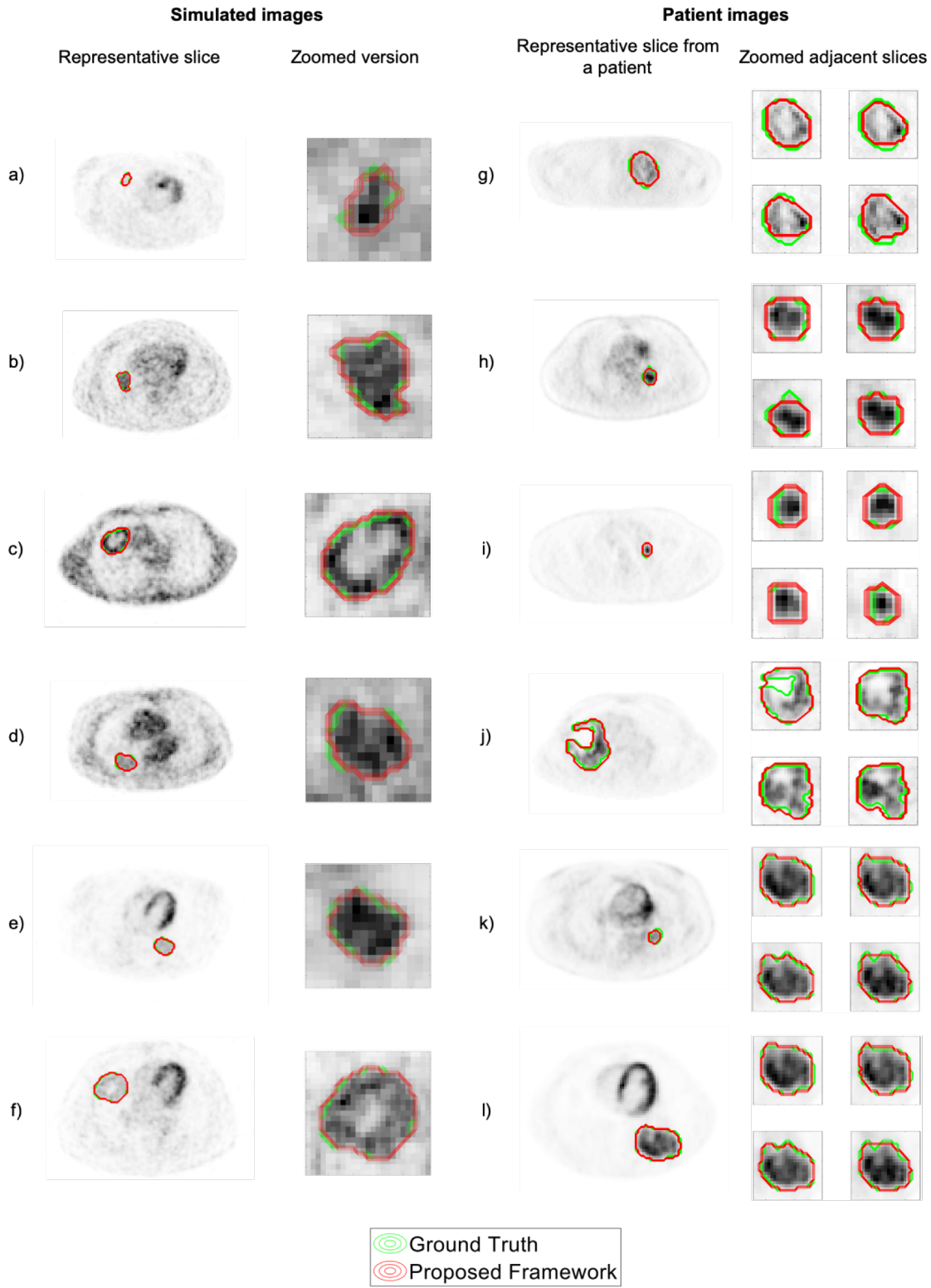

**Figure 3**: Examples of segmented tumors produced by the proposed framework in simulated $^{18}$F-fluorodeoxyglucose (FDG)-PET images where the ground-truth tumor boundaries were known (a-f). Examples of segmented tumors produced by the proposed framework of patient FDG-PET images where manual segmentations were used as ground-truth (g-l). A full representative slice and several adjacent slices, in a zoomed view, are shown per example (g-l).

The kernel distributions of each tumor descriptor were sampled to generate simulated tumors. Intra-tumor heterogeneity was simulated by incorporating unimodal variability in intensity values within the tumor and, for some tumors, by modeling the intensity distribution as a mixture model. For example, tumor cores assigned a lower intensity than the corresponding rim modeled a necrotic tumor. Examples of simulated tumors are shown in Figs. 3a-f.

Table 2: Architecture of the modified U-net.

|  | Layer Type | Filter Size | Stride | # of filters | Input Size | Output Size |
|---|---|---|---|---|---|---|
| Layer 1 | Conv. | 3 × 3 | 1 × 1 | 16 | 128 × 128 × 1 | 128 × 128 × 16 |
| Layer 2 | Conv. | 3 × 3 | 1 × 1 | 16 | 128 × 128 × 16 | 128 × 128 × 16 |
| Layer 3 | Max-pool | 2 × 2 | 2 × 2 | - | 128 × 128 × 16 | 64 × 64 × 16 |
| Layer 4 | Conv. | 3 × 3 | 1 × 1 | 32 | 64 × 64 × 16 | 64 × 64 × 32 |
| Layer 5 | Conv. | 3 × 3 | 1 × 1 | 32 | 64 × 64 × 32 | 64 × 64 × 32 |
| Layer 6 | Max-pool | 2 × 2 | 2 × 2 | - | 64 × 64 × 32 | 32 × 32 × 32 |
| Layer 7 | Conv. | 3 × 3 | 1 × 1 | 64 | 32 × 32 × 32 | 32 × 32 × 64 |
| Layer 8 | Conv. | 3 × 3 | 1 × 1 | 64 | 32 × 32 × 64 | 32 × 32 × 64 |
| Layer 9 | Conv. | 3 × 3 | 1 × 1 | 64 | 32 × 32 × 64 | 32 × 32 × 64 |
| Layer 10 | Max-pool | 2 × 2 | 2 × 2 | - | 32 × 32 × 64 | 16 × 16 × 64 |
| Layer 11 | Transposed Conv. | 2 × 2 | 2 × 2 | 64 | 16 × 16 × 64 | 32 × 32 × 64 |
| Layer 11 (a) | Skip Connection (add output of Layer 9) | - | - | - | 32 × 32 × 64 | 32 × 32 × 64 |
| Layer 12 | Conv. | 3 × 3 | 1 × 1 | 64 | 32 × 32 × 64 | 32 × 32 × 64 |
| Layer 13 | Conv. | 3 × 3 | 1 × 1 | 64 | 32 × 32 × 64 | 32 × 32 × 64 |
| Layer 14 | Conv. | 3 × 3 | 1 × 1 | 64 | 32 × 32 × 64 | 32 × 32 × 64 |
| Layer 15 | Transposed Conv. | 2 × 2 | 2 × 2 | 32 | 32 × 32 × 64 | 64 × 64 × 32 |
| Layer 15 (a) | Skip Connection (add output of Layer 5) | - | - | - | 64 × 64 × 32 | 64 × 64 × 32 |
| Layer 16 | Conv. | 3 × 3 | 1 × 1 | 32 | 64 × 64 × 32 | 64 × 64 × 32 |
| Layer 17 | Conv. | 3 × 3 | 1 × 1 | 32 | 64 × 64 × 32 | 64 × 64 × 32 |
| Layer 18 | Transposed Conv. | 2 × 2 | 2 × 2 | 16 | 64 × 64 × 32 | 128 × 128 × 16 |
| Layer 18 (a) | Skip Connection (add output of Layer 2) | - | - | - | 128 × 128 × 16 | 128 × 128 × 16 |
| Layer 19 | Conv. | 3 × 3 | 1 × 1 | 16 | 128 × 128 × 16 | 128 × 128 × 16 |
| Layer 20 | Conv. | 3 × 3 | 1 × 1 | 2 | 128 × 128 × 16 | 128 × 128 × 2 |
| Layer 21 | Softmax | - | - | - | 128 × 128 × 2 | 128 × 128 × 2 |
| Output | Argmax | - | - | - | 128 × 128 × 2 | 128 × 128 × 1 |

For simulated tumors, the ground-truth tumor boundaries were known. Since the ground-truth for the image background need not be known, we used multiple existing patient images from the training set as templates to provide a realistic tumor background and account for inter-patient variability. For each simulated tumor slice, an FDG-PET patient image slice containing lung but not tumor was selected as background. The tumor locations placed in the patient background image slices were first manually selected such that the simulated tumors would appear at visually realistic locations within the lung. The simulated tumors were generated and randomly placed at the manually selected seed locations within the lung region of the patient background slices. The tumor orientation was determined by applying a random rigid rotation to the tumor. Similar to Ma et al. (Ma *et al* 2017), projection data corresponding to the patient image and simulated tumor slices were generated by simulating a PET system modeling the major image-degrading processes in PET such as detector blurring with a 5 mm full-width-at-half-maximum (FWHM) Gaussian blur and noise at clinical count levels. These data were added in projection space to incorporate the impact of image reconstruction on the tumor appearance and noise texture. The projection data were reconstructed using the 2D ordered subset expectation-maximization algorithm (Hudson and Larkin 1994) with 16 subsets and 3 iterations to yield a large number of simulated images for different phantoms. These reconstruction parameters yielded the most visually realistic images. Another reason for this choice of reconstruction parameters was that 16 subsets with 3 iterations is roughly equivalent to 48 maximum likelihood expectation-maximization iterations, which was approximately equivalent to the number of iterations used to generate the patient images (Table 1). Realism of the generated images was evaluated visually by a board-certified radiologist. The realism of images generated by such an approach has also been evaluated in previous studies (Ma *et al* 2017).

*2.1.2 Module 2 – Training mU-net with simulated data.* The core of the proposed framework was a modified U-net (mU-net) with an encoder-decoder architecture (Fig. 2). The encoder network learns spatially local features from the input data through a series of convolutional layers (Goodfellow et al 2016). Each convolutional layer learns feature maps from the previous layer by performing convolution of the input with a filter bank. After each convolutional layer in the network, a leaky rectified linear unit (ReLU) activation function is applied (Maas et al 2013). The ReLU has been shown to help alleviate the vanishing gradient problem (Maas et al 2013). Max-pooling layers were applied following the convolutional layers in the encoder network to condense meaningful features (Goodfellow et al 2016). The decoder network up-sampled the output of the encoder network via transposed convolutional layers. The transposed convolutional layers performed a learned up-sampling of feature maps by reconstructing the spatial resolution of previous layers prior to the pooling layers. The output of the decoder network was then mapped to a tumor mask in the decoder network. The decoder output was fed into a softmax layer, which performed a pixel-wise tumor classification. The mU-net was trained by minimizing a class-weighted cross-entropy loss function quantifying the error between the predicted and true tumor mask via the Adam optimization algorithm (Pereira et al 2016, Kingma and Ba 2014). A detailed description of the network architecture is given below in Table 2.

There were some differences between the implementation of the (mU-net) and the implementation from the original U-net paper (Ronneberger *et al* 2015). The original U-net implementation does not include the use of dropout whereas the mU-net included dropout after each convolutional layer. Dropout is a well-established regularization technique to prevent overfitting in deep neural networks where neurons and their connections are randomly dropped during training (Srivastava *et al* 2014). Using dropout during training resulted in a stable decrease in the training loss to prevent overfitting and higher segmentation accuracy compared to training the network without dropout. The original U-net has four blocks of expanding and contracting layers whereas our model had three blocks of expanding and contracting layers. Our mU-net architecture used a relatively small number of feature maps per layer (Table 2) compared to the original U-net, since we found that increasing the number of feature maps did not result in substantial gains in segmentation accuracy. The lowest resolution in the original U-net is $32 \times 32$ pixels whereas the lowest resolution in our model was $16 \times 16$ pixels. The mU-net automatically extracted important local contextual and global localization features in the encoder and decoder paths, respectively (Ronneberger *et al* 2015). These features were combined through skip connections similar to the U-net architecture (Ronneberger *et al* 2015). These skip connections allow features learned at the beginning of the network to feed into the later layers and allow the network to learn more complex features. However, instead of feature map concatenation in the skip connections as in the original U-net, our mU-net used element-wise addition between the output of layers in the encoder network to downstream layers in the decoder network. The use of skip connections with element-wise addition in the mU-net stabilized training and helped improve performance. Finally, the original U-net was developed for segmentation in biomedical microscopy images (Ronneberger *et al* 2015) whereas the modified U-net was optimized and fine-tuned for PET images.

The mU-net architecture and hyperparameters were optimized via 5-fold cross-validation on the simulated images (Fig. 1b). A grid search was used for the hyperparameter optimization where the general range for each hyperparameter sweep spanned several orders of magnitude. Hyperparameters included the value for the α parameter used for the Leaky ReLU activation function, the dropout probability, the initialization value of the bias term for all weights in the network, and the class-weighting on the cross-entropy loss function. It was found during the cross-validation process that initializing the network weights by the Glorot initialization procedure helped to address the problem of vanishing or exploding backpropagated gradients (Glorot and Bengio 2010). Additionally, since there were relatively few tumor pixels compared to background pixels, class balancing on the cross-entropy loss function was used by weighting the loss more heavily for tumor pixels (Pereira *et al* 2016). The detailed list of hyperparameters are shown in Table 3. After the hyperparameter selection, the final mU-net architecture, which had been trained only on subsets of the training data via the 5-fold cross-validation, was then trained with the entire training set from scratch. The hyperparameters that performed best on the training set during the 5-fold cross-validation were used to train the network during this step.

Table 3: Hyperparameters of proposed framework.

|                | Hyperparameter       | Value  |
|----------------|----------------------|--------|
| Initialization | weights              | Xavier |
|                | bias                 | 0.03   |
| Leaky ReLU     | α                    | 0.01   |
| Dropout        | dropout probability  | 0.1    |
| Loss function  | class weighting      | 2:1    |
| Training       | epochs               | 200    |
|                | batch size           | 25     |
| Fine-tuning    | epochs               | 200    |
|                | batch size           | 10     |

*2.1.3 Module 3 – Fine-tuning with a small amount of clinical data.* The objective of this module was to fine-tune the mU-net with patient data to learn tumor features that may have been missed in simulated tumors. The pre-trained network was fine-tuned using a small-sized clinical dataset (Fig. 1c) where the weights of the pre-trained network were used to initialize training of the fine-tuned network on patient data. The approach was similar to the fine-tuning process used in certain transfer learning-based approaches (Van Opbroek *et al* 2014). Primary tumors in the dataset were segmented by a board-certified radiologist with 4 and 11 years of experience in nuclear-medicine and diagnostic radiology, respectively. Manual segmentation was performed on a per-slice basis on 2D transaxial image slices. These manual segmentations were considered as ground-truth tumor boundaries. An overview of the proposed framework is shown in Fig. 1d.

*2.2 Evaluating the proposed framework*

The framework was comprehensively evaluated via multiple experiments with independent training and test sets (Fig. 1e). The framework's accuracy on accurately quantifying tumor segmentation and localization in image slices was quantified using Dice similarity coefficient (DSC), Jaccard similarity coefficient (JSC), true positive fraction, true negative fraction, Hausdorff distance (HD) (Foster *et al* 2014, Soufi *et al* 2016), and tumor localization (TL) accuracy. DSC and JSC measure the spatial overlap between the delineated and true tumor masks. Higher values indicate more accurate segmentation (Foster *et al* 2014). DSC values of 0.7 indicate high segmentation accuracy as mentioned in Zijdenbos et al. (Zijdenbos *et al* 1994). True positive fraction and true negative fraction denote the fraction of correctly identified tumor pixels and background pixels, respectively. TL accuracy quantifies the fraction of times there was any overlap between the predicted and ground-truth segmentations. The HD quantifies shape similarity between the delineated and true tumor boundaries (Foster *et al* 2014, Soufi *et al* 2016), with lower values indicating higher shape similarity. HD was computed only for correct TL to quantify shape similarity without the effect of localization errors. All other metrics were computed generally irrespective of correct TL, quantifying performance on the joint tumor localization-segmentation task. DSC values were computed both generally and for correct TL. All values are reported as mean (95% confidence intervals (CIs)). Statistical significance was determined using a paired sample *t*-test where a *p*-value < 0.01 was used to infer a statistically significant difference.

*2.2.1 Evaluating accuracy and comparing to other methods using patient and simulated data.* The framework was quantitatively compared to semi-automated segmentation methods, including commonly used thresholding-based approaches (Foster *et al* 2014, Mena *et al* 2017a, Shah *et al* 2012, Mena *et al* 2017b, Sridhar *et al* 2014), the active contour-based snakes method (Kass *et al* 1988), and the Markov random fields-Gaussian mixture model-based clustering technique (Jha *et al* 2010, Layer *et al* 2015). Three thresholding approaches using fixed intensity thresholds of 30%, 40%, and 50% of $SUV_{max}$ were considered (Sridhar *et al* 2014). The MRF-GMM method was optimized to yield the best overlap with ground truth on the basis of DSC. A range of values for beta, a parameter in the MRF-GMM method, were tested on the simulated data and the beta value that resulted in the best DSC was selected. The semi-automated techniques were provided the tumor location as user input through a seed pixel or ROI. In contrast, the proposed automated framework had to both localize and segment the tumor, a more challenging task.

For the thresholding-based approaches, a procedure similar to (Sridhar *et al* 2014) was followed. A circular ROI around the tumor was provided to all thresholding-based approaches. The tumor was segmented by applying a threshold of 30%, 40%, and 50% SUVmax within the circular ROI using region growing (Day *et al* 2009) starting from the SUVmax pixel. For the active contour-based method, a circular ROI around the tumor was provided, and the tumor contour was generated by performing the active contour segmentation procedure to the ROI. For the MRF-GMM-based clustering method, a procedure similar to (Jha *et al* 2010) was followed.

Evaluation was done with simulated and patient data (Fig. 1e). Evaluation with simulated data assessed the framework's performance with known ground truth. 10,000 image slices were simulated based on data from 113 randomly selected patient images (Module 1 of framework). The network was trained on this simulated data as in Fig. 1b (Module 2). Next, 2,000 completely independent slices were generated and used to test the network performance on simulated data.

Evaluation with patient data assessed the performance on clinical tumor images with manual segmentations as surrogate ground-truth. The framework pre-trained with simulated data was fine-tuned with the 113 patient images used to generate the simulated data (Module 3) and evaluated with the remaining 47 patient images.

*2.2.2 Evaluating generalizability of the proposed framework.* The generalizability of DL-based approaches to data acquired from different scanners is highly important otherwise the DL approach would have to be retrained using data acquired from every scanner, making the approach impractical (Chang *et al* 2018).

To evaluate our framework's generalizability, we used the fact that patient images in our clinical dataset were acquired using five different scanners. Two experiments were performed. In the first experiment, the 104 patient images acquired from the Discovery LS scanner were used for simulations, pre-training and fine-tuning. The framework's performance was tested on the 56 patient images from all other scanners. This experiment evaluated the framework's generalizability when trained on images from one scanner and tested on images from other multiple scanners.

In the second experiment, 56 patient images from the Discovery RX, HR, ST, and STE scanners were used for simulations, pre-training, and fine-tuning. Testing was with 104 patient images from the Discovery LS scanner, thus evaluating the framework's generalizability when trained on images from multiple scanners and tested on images from another scanner.

*2.2.3 Evaluating the efficacy of the framework in reducing number of clinical training images.* For this purpose, the framework's performance was compared to that of a mU-net trained only on clinical data in two experiments.

In the first experiment, the framework was pre-trained with simulated images based on 104 patient images acquired by the Discovery LS scanner. The size of the clinical training set, which consisted patient images from the Discovery LS scanner, was varied from 1, 5, 25, 50, 75, and 104 patients. Both the framework and the mU-net trained only on clinical data were evaluated on a test set of 56 patient images from all other scanners.

In the second experiment, the framework was pre-trained with simulated images based on the 56 patient images from all other scanners. The size of the clinical training set, which consisted of patient images from all other scanners, was varied from 1, 5, 20, 30, 40, and 56 patients. The testing was with the 104 patient images from the Discovery LS scanner. The performance was quantitatively compared on the tasks of segmentation and tumor localization in both experiments.

*2.2.4 Evaluating sensitivity of the framework to PVEs.* For this purpose, experiments were performed with the test set of 2,000 simulated image slices as defined in Section 2.2.1. Simulated images were used since the ground-truth tumor masks for these images were known. These ground-truth masks were blurred by applying a rectangular filter that incorporated the resolution degradation due to the imaging system and reconstruction, yielding a tumor boundary now affected by PVEs.

These PVE-affected tumor boundaries and the tumor boundaries estimated by the proposed framework were compared to the ground-truth on the basis of DSC and the ratio between the measured and the true tumor areas in the slices (referred to as %area) (De Bernardi *et al* 2009). A %area of 100% indicates perfect tumor-area prediction, while greater or lesser than 100% indicates overestimation and underestimation of the tumor area, respectively. Only cases where the network correctly localized the tumor were considered in order to specifically study sensitivity to PVEs.

*2.3 Implementation details*

The network architecture and training were implemented in Python 3.4.5, TensorFlow 1.6.0, and Keras 2.1.5. Experiments were run on an NVIDIA Tesla K40 GPU and a Linux CentOS 5.10 operating system. A detailed list of the hardware and software used to implement the network architecture are given in Table 4.

Table 4: Hardware and software platform.

| Graphics Processing Unit (GPU) | Model | NVIDIA Tesla K40 |
|---|---|---|
| Operating System | Linux | CentOS 5.10 |
| Deep Learning (DL) Platform | Programming Language | Python 3.4.5 |
| | Open-source DL libraries | TensorFlow 1.6.0, Keras 2.1.5 |

## 3. Results

### 3.1 Evaluating accuracy and comparing to other methods using patient and simulated data

The proposed framework quantitatively outperformed all other semi-automated methods on the basis of DSC, JSC, and HD ($p$-values<0.001) for both simulated and patient images (Figs. 4a-b and Table 5). The framework yielded a DSC of 0.87 (95% CI: 0.86, 0.88) and 0.73 (95% CI: 0.71, 0.76) on simulated and patient images, respectively, indicating reliable segmentation performance (Zijdenbos *et al* 1994). Further, the framework yielded a DSC of 0.91 (95% CI: 0.91, 0.92) and 0.84 (95% CI: 0.83, 0.85) for simulated and patient images, respectively, when TL was correct.

Qualitatively, there was a good match between tumor segmentations obtained with the proposed framework and both the ground-truth for simulated images and the radiologist-defined segmentation for patient images (Fig. 3). The framework segmented tumors that had substantial intra-tumor heterogeneity (Figs. 3c, f and g), were surrounded by regions with high uptake (Figs. 3b, d and h), were relatively small (Figs. 3a and i), had convex shapes (Fig. 3j), and were near the heart region (Figs. 3e, f, k, l). The smallest segmented tumor axial cross-section was 1.83 cm$^2$ in area.

### 3.2 Evaluating generalizability of the proposed framework

The proposed framework provided reliable segmentation and yielded a DSC of 0.74 (95% CI: 0.71, 0.76) and 0.71 (95% CI: 0.69, 0.73) on the test sets of the first and second generalization experiments described in Section 2.2.2, respectively (Figs. 4c-d and Table 6), outperforming other semi-automated techniques on the basis of DSC, JSC, and HD ($p$-value<0.001). When TL was correct, the framework yielded a DSC of 0.85 (95% CI: 0.84, 0.86) and 0.82 (95% CI: 0.81, 0.84) on the test sets of the first and second generalization experiments described in Section 2.2.2, respectively. This provided evidence that the framework generalized across different PET scanners.

### 3.3 Evaluating efficacy of the framework in reducing number of clinical training images

The proposed framework outperformed the mU-net trained only on clinical data on the basis of DSC and localization accuracy (Figs. 5a-d) for all training sizes (*p*-values<0.001). Representative examples in Figs. 5e-f demonstrate this further. When trained on patient images acquired by the Discovery LS scanner, the proposed framework yielded a DSC of 0.68 (95% CI: 0.66, 0.71) even when trained on 25 patient images (Fig. 5a) and a localization accuracy of 74% (95% CI: 71%, 77%) when trained with just one patient image (Fig. 5b). When TL was correct, the proposed framework yielded a DSC of 0.79 (95% CI:

**Table 5**: Comparing the proposed framework and other methods on simulated images and patient images using the procedure in Section 2.2.1.

| Simulated Images | | | | | | |
|---|---|---|---|---|---|---|
| Segm. Methods | mU-net | MRF-GMM | Snakes | 30% $SUV_{max}$ | 40% $SUV_{max}$ | 50% $SUV_{max}$ |
| DSC | 0.87 (0.86, 0.88) | 0.61 (0.60, 0.63) | 0.58 (0.57, 0.60) | 0.58 (0.56, 0.59) | 0.63 (0.61, 0.64) | 0.57 (0.56, 0.59) |
| DSC with correct TL | 0.91 (0.91, 0.92) | 0.61 (0.60, 0.63) | 0.58 (0.57, 0.60) | 0.58 (0.56, 0.59) | 0.63 (0.61, 0.64) | 0.57 (0.56, 0.59) |
| JSC | 0.81 (0.80, 0.82) | 0.50 (0.48, 0.51) | 0.47 (0.45, 0.48) | 0.48 (0.47, 0.50) | 0.55 (0.54, 0.57) | 0.49 (0.47, 0.50) |
| TPF | 0.90 (0.89, 0.91) | 0.86 (0.86, 0.87) | 0.88 (0.88, 0.89) | 0.80 (0.79, 0.82) | 0.68 (0.66, 0.70) | 0.53 (0.51, 0.55) |
| TNF | 1.00 (1.00, 1.00) | 0.99 (0.98, 0.99) | 0.99 (0.99, 0.99) | 0.99 (0.99, 0.99) | 1.00 (1.00, 1.00) | 1.00 (1.00, 1.00) |
| HD | 1.45 (1.39, 1.51) | 6.05 (5.75, 6.34) | 11.61 (11.29, 11.93) | 7.76 (7.44, 8.07) | 6.52 (6.17, 6.86) | 6.78 (6.43, 7.14) |
| Patient Images | | | | | | |
| Segm. Methods | mU-net | MRF-GMM | Snakes | 30% $SUV_{max}$ | 40% $SUV_{max}$ | 50% $SUV_{max}$ |
| DSC | 0.73 (0.71, 0.76) | 0.68 (0.66, 0.70) | 0.67 (0.65, 0.68) | 0.66 (0.64, 0.68) | 0.60 (0.58, 0.62) | 0.50 (0.48, 0.52) |
| DSC with correct TL | 0.84 (0.83, 0.85) | 0.68 (0.66, 0.70) | 0.67 (0.65, 0.68) | 0.66 (0.64, 0.68) | 0.60 (0.58, 0.62) | 0.50 (0.48, 0.52) |
| JSC | 0.65 (0.63, 0.68) | 0.55 (0.53, 0.57) | 0.52 (0.51, 0.53) | 0.53 (0.51, 0.55) | 0.46 (0.44, 0.47) | 0.36 (0.35, 0.38) |
| TPF | 0.76 (0.73, 0.79) | 0.86 (0.84, 0.88) | 0.68 (0.67, 0.70) | 0.63 (0.61, 0.65) | 0.50 (0.48, 0.52) | 0.38 (0.36, 0.39) |
| TNF | 1.00 (1.00, 1.00) | 0.99 (0.98, 0.99) | 1.00 (1.00, 1.00) | 1.00 (1.00, 1.00) | 1.00 (1.00, 1.00) | 1.00 (1.00, 1.00) |
| HD | 3.25 (2.92, 3.58) | 5.40 (4.88, 5.92) | 7.14 (6.62, 7.67) | 5.50 (4.99, 6.01) | 5.63 (5.12, 6.14) | 6.11 (5.60, 6.63) |

Note – Data in parentheses are 95% confidence intervals. DSC: Dice similarity coefficient, HD: Hausdorff distance, JSC: Jaccard similarity coefficient, MRF-GMM: Markov random fields-Gaussian mixture model, mU-net: modified U-net, TL: tumor localization, TNF: true negative fraction, TPF: true positive fraction. Sample sizes were 1,916 and 486 for DSC with correct TL and HD of mU-net for simulated and patient images, respectively. Sample sizes were 2,000 and 557 for all other metrics and segmentation methods for simulated and patient images, respectively.

0.77, 0.80) when trained on 25 patient images (Fig. 5a). Similarly, when trained on patient images from all other scanners, the proposed framework yielded a DSC of 0.70 (95% CI: 0.68, 0.71) when trained on 30 patient images (Fig. 5c) and a localization accuracy of 72% (95% CI: 69%, 75%) when trained with just one patient image (Fig. 5d).

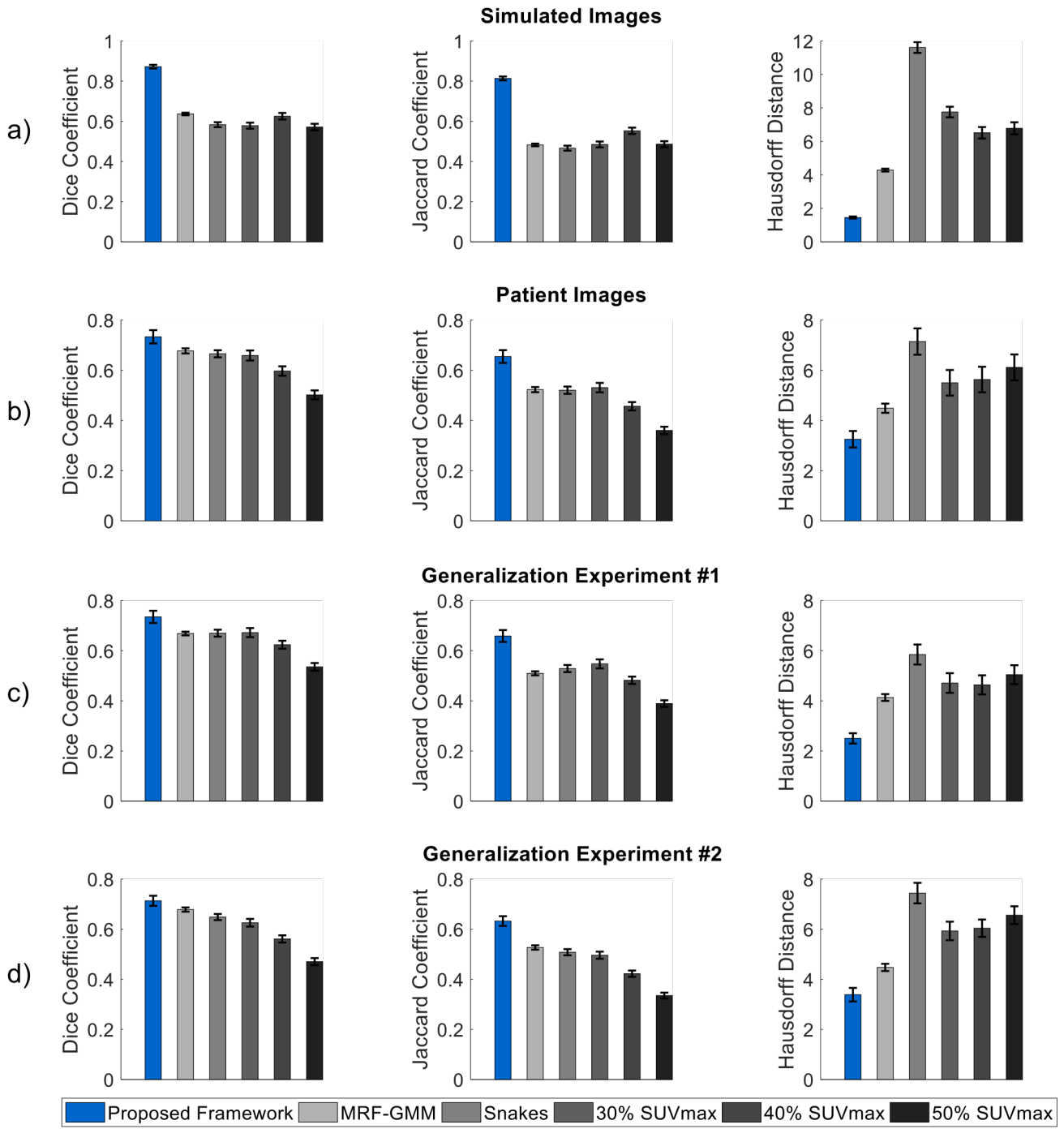

**Figure 4**: Comparing the proposed framework and other semi-automated techniques for simulated images (a) and patient images (b) using the procedure in Section 2.2.1. Evaluating the generalizability of the proposed framework across multiple PET scanners when trained on images acquired by a single PET scanner (c) and across a single PET scanner when trained on images acquired by multiple PET scanners (d) using the procedure in Section 2.2.2. Sample sizes were 2,000 (a), 557 (b), 662 (c), and 1001 (d). Error bars represent 95% confidence intervals. MRF-GMM: Markov random fields-Gaussian mixture model.

*3.4 Evaluating sensitivity of the framework to PVEs*

The network correctly localized the tumor in 1,916 of the 2,000 images (95.8%) from the test set of simulated images. These cases that were correctly localized were only considered here to study sensitivity to PVEs. Representative examples comparing the predicted tumor boundaries by the proposed framework to the PVE-affected tumor boundaries are shown in Figs. 6a-d. The proposed framework yielded a DSC of 0.91 (95% CI: 0.91, 0.92) while the PVE-affected tumor boundaries yielded a DSC of 0.75 (95% CI: 0.74, 0.75) for simulated images. The proposed framework outperformed (*p*-value<0.001) the PVE-affected tumor boundaries on the basis of DSC and %area (Figs. 6e-f) and yielded reliable segmentation and accurate tumor-area prediction for all tumor sizes (Fig. 6).

## 4. Discussion

We proposed a modular automated DL-based framework for tumor segmentation in PET images. The framework accurately localized and delineated primary tumors on FDG-PET images of patients with lung cancer using a small-sized clinical training dataset. The framework generalized across several PET scanners, indicating the features learned by the framework were invariant to scanner differences. Those attributes provide evidence that the framework is robust and can be implemented by various institutions and centers with different PET scanners. Further, the proposed framework outperformed other semi-automated methods for both simulated and patient images.

Visually, the proposed framework successfully localized the primary lung tumor even in cases where multiple high-uptake regions were present within the same image (e.g. heart, mediastinum, lymph nodes, or secondary metastatic deposits) (Figs. 3 and 5e). Concurrently, there were few cases where the DL approach could not localize the tumor correctly (Fig. 5g). However, the mU-net trained only on clinical images also failed in those cases (Fig. 5h). The localization accuracy of the proposed framework was generally higher than 80%, and up to 91% when data from 104 patients were used for training (Fig. 5b). To address cases of inaccuracy, one solution would be to display the segmented-tumor output to a radiologist for approval or refinement and integrate this feedback with a reinforcement-learning approach, similar to that in (Wang *et al* 2018).

**Table 6**: Comparing the proposed framework and other methods on patient images using the procedure in Section 2.2.2.

| Segm. Methods | mU-net | MRF-GMM | Snakes | 30% $SUV_{max}$ | 40% $SUV_{max}$ | 50% $SUV_{max}$ |
|---|---|---|---|---|---|---|
| **Generalizability Experiment #1** | | | | | | |
| DSC | 0.74 (0.71, 0.76) | 0.68 (0.67, 0.70) | 0.67 (0.66, 0.68) | 0.67 (0.65, 0.69) | 0.62 (0.61, 0.64) | 0.54 (0.52, 0.55) |
| DSC with correct TL | 0.85 (0.84, 0.86) | 0.68 (0.67, 0.70) | 0.67 (0.66, 0.68) | 0.67 (0.65, 0.69) | 0.62 (0.61, 0.64) | 0.54 (0.52, 0.55) |
| JSC | 0.66 (0.64, 0.68) | 0.55 (0.53, 0.57) | 0.53 (0.51, 0.54) | 0.55 (0.53, 0.57) | 0.48 (0.47, 0.50) | 0.39 (0.38, 0.40) |
| TPF | 0.76 (0.74, 0.79) | 0.83 (0.81, 0.84) | 0.71 (0.70, 0.72) | 0.71 (0.69, 0.73) | 0.56 (0.55, 0.58) | 0.42 (0.41, 0.44) |
| TNF | 1.00 (1.00, 1.00) | 1.00 (0.99, 1.00) | 1.00 (1.00, 1.00) | 1.00 (1.00, 1.00) | 1.00 (1.00, 1.00) | 1.00 (1.00, 1.00) |
| HD | 2.50 (2.30, 2.71) | 4.17 (4.00, 4.44) | 5.85 (5.45, 6.25) | 4.71 (4.32, 5.10) | 4.64 (4.26, 5.02) | 5.04 (4.67, 5.42) |
| **Generalizability Experiment #2** | | | | | | |
| DSC | 0.71 (0.69, 0.73) | 0.66 (0.65, 0.67) | 0.65 (0.64, 0.66) | 0.63 (0.61, 0.64) | 0.56 (0.55, 0.58) | 0.47 (0.46, 0.48) |
| DSC with correct TL | 0.82 (0.81, 0.84) | 0.66 (0.65, 0.67) | 0.65 (0.64, 0.66) | 0.63 (0.61, 0.64) | 0.56 (0.55, 0.58) | 0.47 (0.46, 0.48) |
| JSC | 0.63 (0.61, 0.65) | 0.53 (0.52, 0.55) | 0.51 (0.50, 0.52) | 0.50 (0.48, 0.51) | 0.42 (0.41, 0.43) | 0.34 (0.32, 0.35) |
| TPF | 0.72 (0.70, 0.74) | 0.83 (0.82, 0.85) | 0.69 (0.68, 0.70) | 0.61 (0.60, 0.63) | 0.48 (0.46, 0.49) | 0.36 (0.34, 0.37) |
| TNF | 1.00 (1.00, 1.00) | 0.99 (0.98, 0.99) | 0.99 (0.99, 1.00) | 1.00 (1.00, 1.00) | 1.00 (1.00, 1.00) | 1.00 (1.00, 1.00) |
| HD | 3.38 (3.11, 3.66) | 5.79 (5.40, 6.19) | 7.44 (7.03, 7.85) | 5.93 (5.56, 6.30) | 6.04 (5.69, 6.39) | 6.56 (6.20, 6.91) |

Note – Data in parentheses are 95% confidence intervals. DSC: Dice similarity coefficient, HD: Hausdorff distance, JSC: Jaccard similarity coefficient, MRF-GMM: Markov random fields-Gaussian mixture model, mU-net: modified U-net, TL: tumor localization, TNF: true negative fraction, TPF: true positive fraction. Sample sizes were 574 and 867 for DSC with correct TL and HD of mU-net for experiments #1 and #2, respectively. Sample sizes were 662 and 1001 for all other metrics and segmentation methods for experiments #1 and #2, respectively.

Experiments with simulated data demonstrated that the proposed framework was relatively insensitive to PVEs (Fig. 6). The proposed framework successfully segmented relatively small tumors in patient images, despite the presence of PVEs. These results demonstrate the applicability of the proposed framework in modalities with limited resolution and lack of ground-truth, such as SPECT and optical imaging modalities.

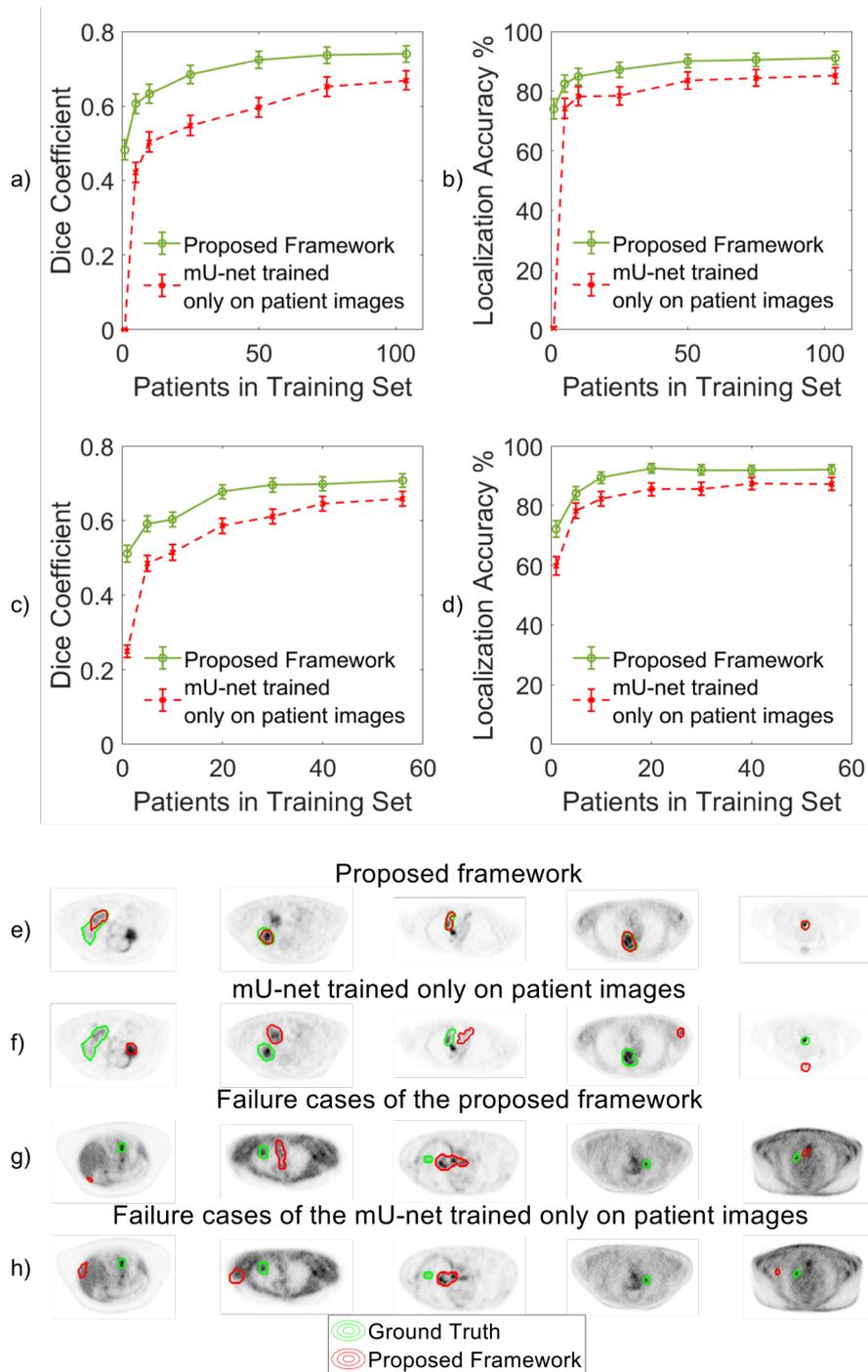

**Figure 5**: Comparing the proposed framework to the modified U-net (mU-net) trained only on clinical images in terms of Dice similarity coefficient and tumor localization accuracy for the various training set sizes with 95% confidence intervals (a-d) using the procedure in Section 2.2.3. Representative examples of segmented tumors by the proposed framework (e) and the mU-net that was trained on clinical images only (f). Each example in (e) and (f) refer to the same image slices. Cases where the proposed framework (g) and the mU-net that was trained on only clinical images (h) failed to localize the tumor are also shown. Similarly, each example in (g) and (h) refer to the same image slices.

The proposed framework uses a new stochastic-KDE and physics-based approach to generate realistic simulated images that address limited availability of clinical training data. This approach allows the generated data to account for patient-population variability and simultaneously account for the imaging physics. Other data-augmentation strategies include transforming the tumor (e.g. translation, rotation, scaling), or changing tumor intensity in the patient images (Litjens *et al* 2017, Pereira *et al* 2016, Shen *et al* 2017). Even for such augmented data, the ground-truth tumor boundary would be the manual segmentations, which suffer from several issues already described. Our attempts at training the mU-net with these strategies were ineffective.

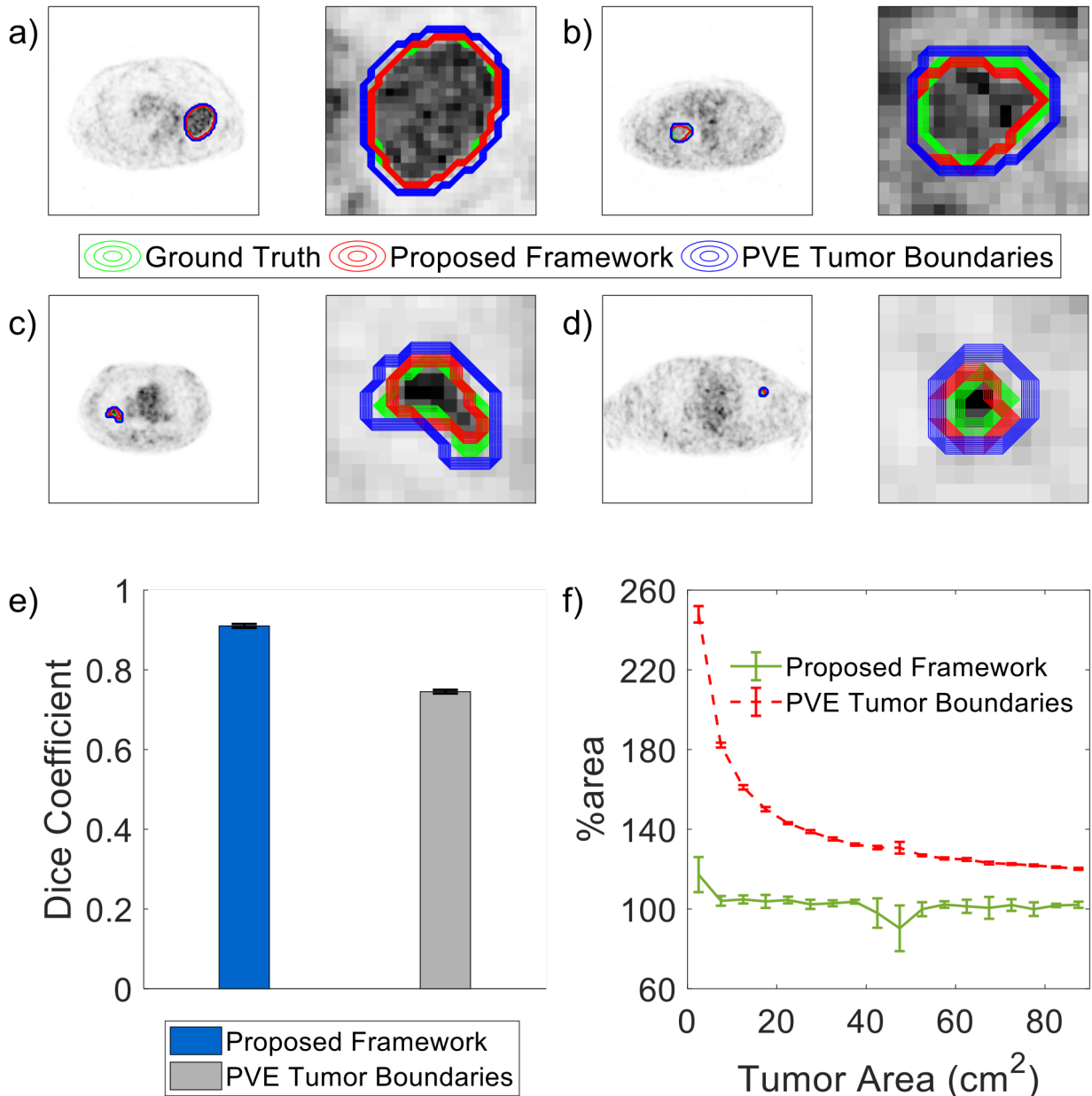

**Figure 6**: Representative examples comparing the predicted tumor boundaries by the proposed framework to the tumor boundaries affected by partial volume effects (PVEs) in simulated images (a-d). Evaluating the ability of the proposed framework to compensate for PVEs on the basis of Dice similarity coefficient (e), and %area (f) using the procedure in Section 2.2.4. A sample size of 1916 simulated images from the test set where the tumor was correctly localized were used (e). The %area was plotted as a function of true tumor area (cm$^2$) to demonstrate the effect of PVEs for different tumor sizes where the tumor sizes were binned with a bin width of 5 cm$^2$ tumor area. The sample sizes for each bin in order of increasing tumor area were 405, 457, 284, 203, 162, 111, 74, 45, 41, 16, 33, 18, 11, 11, 13, 10, 15, and 7, respectively. Error bars represent 95% confidence intervals.

Another strategy would be the use of generative adversarial networks (GANs) (Gong *et al* 2018) trained with clinical data. However, GANs can suffer from stability issues (Creswell *et al* 2018) and do not exploit the known physics of PET imaging unlike the proposed framework.

The framework was developed for cases where only the PET image data is available, as is the case with data acquired using PET-only systems or scans. However, with PET/CT and PET/MRI scanners, multimodality imaging can assist with segmentation. Notably, the study done in (Zhao *et al* 2018) developed a multi-modal CNN-based method for tumor co-segmentation in PET/CT images. Our framework was also not developed for PET scanners that implement time-of-flight (TOF). Recent work on tumor segmentation using TOF PET has demonstrated promise (Blanc-Durand *et al* 2018). Integrating the proposed modular structure with these approaches could improve accuracy and reduce the need for large training data.

Our study has some limitations. The proposed framework performs 2D tumor delineation on PET transaxial image slices rather than on the entire 3D volume and thus assumes that the input image slices have a tumor. This was due to several reasons. Firstly, this was in line with our objective in this manuscript, which was on developing a method that, when given a PET slice with a tumor, localizes the tumor and yields accurate tumor boundaries that are relatively insensitive to the partial volume effects in PET images. In other words, we focused on the tumor localization and delineation task, and not on a 3D segmentation approach that would also require tumor-detection such as in Zhao et al (Zhao *et al* 2018). Next, using all tumor-containing 2D slices as training examples increased the amount of training data. Further, 2D tumor delineation is less computationally expensive. Finally, from a usage perspective, the framework can segment 3D images on a per-slice basis when it is provided transaxial image slices where the primary tumor is present, as in this study. However, fully 3D segmentation could allow the network to learn important 3D information to segment contiguous tumor volumes. Extending the proposed framework to direct 3D segmentation is an important research area. Another limitation of our study is in the simulation framework. The framework generated realistic simulated images where synthetic tumors are generated and randomly placed in the lung region of a patient background slice containing no tumor. While this approach yields visually realistic simulated PET images with lung tumors, this does not capture the biological and clinical information about the tumor location in the clinical dataset. Extending the proposed method to incorporate this information in simulations is an important area of research. Also, the proposed framework only locates and segments a single tumor per image slice. The latter limitation is an outcome of the patient selection criteria, where patients with a second primary malignancy were excluded. Extending this method to find multiple tumors per patient image is another important research area.

## 5. Conclusion

A physics-guided modular DL-framework for automated tumor segmentation was developed and provided accurate delineation of primary lung tumors in FDG-PET images with a small-sized clinical training dataset, generalized across different scanners, and demonstrated ability to segment small tumors. Open-source code for the proposed framework is available here: https://drive.google.com/drive/folders/1H483HB-byS6UiPlf3ip1i52i0EH7R4lK?usp=sharing.

## Acknowledgements

Financial support for this project was provided, in part, by the National Institutes of Health under grant terns P41-EB024495 and R21-EB024647. The researchers thank Dr. Martin Lodge and Martin Stumpf for helpful discussions.

## Appendix.

Table A1: Patient characteristics.

| Characteristic | | percent |
|---|---|---|
| Age | <40 | 1% (2/160) |
| | 40-60 | 43% (69/160) |
| | >60 | 56% (89/160) |
| Sex | Male | 57% (91/160) |
| | Female | 43% (69/160) |
| Race | White | 68% (109/160) |
| | Black | 21% (34/160) |
| | Other | 11% (17/160) |
| Histology | Non-small cell lung cancer | 82% (131/160) |
| | Small cell lung cancer | 16% (26/160) |
| | Mesothelioma | 1% (1/160) |
| | Unknown | 1% (2/160) |
| Smoking history | Positive | 79% (126/160) |
| | Negative | 14% (22/160) |
| | Unknown | 8% (12/160) |
| Stage | I | 11% (18/160) |
| | II | 9% (15/160) |
| | III | 34% (54/160) |
| | IV | 38% (61/160) |
| | Unknown | 8% (12/160) |
| Treatment | Surgery | 3% (4/160) |
| | Chemotherapy | 26% (41/160) |
| | Radiation therapy | 6% (9/160) |
| | Surgery & Chemoradiation | 10% (16/160) |
| | Chemoradiation | 48% (77/160) |
| | Surgery & Chemotherapy | 8% (13/160) |
| Interval between treatment and PET study | 1-8 weeks | 65 % (104/160) |
| | 8-12 weeks | 11% (17/160) |
| | 12-24 weeks | 22% (35/160) |
| | 24-40 weeks | 3% (4/160) |
| Outcome | Alive | 33% (52/160) |

The data used in this study is a subset of previously reported data (Sheikhbahaei *et al* 2016), although the purpose of our study is very different. Details about the patient charactertistics are given in Table A1.